\documentclass[%
aps, %
prl, floatfix, 
twocolumn, superscriptaddress, 10pt, 
a4paper, %
amsfonts, amsmath, amssymb]{revtex4-1}
\usepackage[sort&compress]{natbib}
\usepackage{newtxtext}
\usepackage[varg]{newtxmath}
\usepackage{ascmac}
\usepackage{bm}
\usepackage{color}
\usepackage{graphicx}
\usepackage{hyperref}
\hypersetup{%
    unicode=false, %
    bookmarksnumbered=true, %
    bookmarkstype=toc, %
    breaklinks=true, %
    colorlinks=true, %
    pdfborder={0 0 1}, %
    linkcolor=blue, %
    citecolor=blue, %
    urlcolor=blue, %
    pdftoolbar=true, %
    pdfmenubar=true, %
    pdffitwindow=false, %
    pdfstartview={FitH} %
}
\newcommand{\br}{\bm{r}}

\newcommand{\bp}{\bm{p}}
\newcommand{\bq}{\bm{q}}
\newcommand{\kB}{k_{\mathrm{B}}}
\newcommand{\zi}{i}
\newcommand{\dd}[1]{\mathrm{d} #1\,}

\renewcommand{\Im}{\mathrm{Im}}
\newcommand{\R}{\mathrm{R}}

\begin{document}
%
\title{Magnon Current Generation by Dynamical Distortion}
\date{\today}
\author{Junji Fujimoto}
\email[Email:]{junji@ucas.ac.cn}
\affiliation{Kavli Institute for Theoretical Sciences, University of Chinese Academy of Sciences, Beijing, 100190, China}

\author{Mamoru Matsuo}
\affiliation{Kavli Institute for Theoretical Sciences, University of Chinese Academy of Sciences, Beijing, 100190, China}
\affiliation{RIKEN Center for Emergent Matter Science, Wako, Saitama 351-0198, Japan}
\affiliation{Advanced Science Research Center, Japan Atomic Energy Agency, Tokai, 319-1195, Japan}
\affiliation{CAS Center for Excellence in Topological Quantum Computation, University of Chinese Academy of Sciences, Beijing 100190, China}
\begin{abstract}
    The interaction between spin and nanomechanical degrees of freedom attracts interest from the viewpoint of basic science and device applications.
    We study the magnon current induced by the torsional oscillation of ferromagnetic nanomechanical cantilever.
    We find that a finite Dzyaloshinskii-Moriya~(DM) interaction emerges by the torsional oscillation, which is described by the spin gauge field, and the DM interaction leads to the detectably-large magnon current with frequency same as that of the torsional oscillation.
    Our theory paves the way for studying torsional spin-nanomechanical phenomena by using the spin gauge field.
\end{abstract}
\maketitle

The interplay between magnetism and mechanics has a long history, in which the magnetomechanical effect named the Einstein-de Haas effect~\cite{richardson1908,einstein1915}, as well as the inverse effect~\cite{barnett1909,barnett1915}, still attracts interest~\cite{wallis2006,ganzhorn2016,dornes2019,imai2018,imai2019}.
Meanwhile, successive developments in modern technology allow us to decrease the size of mechanical systems down to the nanoscale with high accuracy~\cite{cleland2003,ekinci2005}, which is called nanoelectromechanical system~(NEMS), where quantum mechanics plays an essential role~\cite{oconnell2010,chan2011}.
In addition to the interest from the viewpoint of fundamental physics, the application of NEMS diverges into many branches, such as atomic mass sensing~\cite{jensen2008}, biological imaging~\cite{kalinin2007}, and quantum measurement~\cite{lahaye2009}.
With the developments, the mutual interaction between spin and nanomechanical degrees of freedom has drawn much attention; the electron spin flip observed as a nanomechanical torque~\cite{mohanty2004,zolfagharkhani2008}.
Among these, theoretical proposals for spintronic applications by using nanomechanical motion are also presented, such as magnetization reversal~\cite{kovalev2005}, spin polarization of electric current~\cite{kovalev2008}, and detection of spin Hall effect~\cite{boales2016}, but most of them are for electronic nanomechanical systems.

More recently, a \textit{ferromagnetic insulating} mechanical cantilever of submicron scale was first fabricated~\cite{seo2017}, and by using such a cantilever, a thermally-induced magnetomechanical effect was observed by Harii \textit{et al}.~\cite{harii2019}.
In the experiment on the yttrium-iron-garnet~(YIG) cantilever, the spin wave propagation excited by spin Seebeck effect~\cite{uchida2008} affects the mechanical oscillation of the cantilever, where the authors observe the effect as the resonant frequency modulation of the oscillation.
This experiment is distinguished in a sense that the effect arises in the absence of conduction electron, which means that ferromagnetic spins directly couple to the nanomechanical motion.
Here, one may expect the inverse phenomenon of the effect: nanomechanical motion induces spin wave propagation, which is an interesting effect as fundamental physics.
 It may also stimulate device applications, \textit{e.g.}, nanomechanical spin-wave generator.
However, no one yet shows such a phenomenon, even in theory.

\begin{figure}[b]
	\centering
	\includegraphics[width=\linewidth]{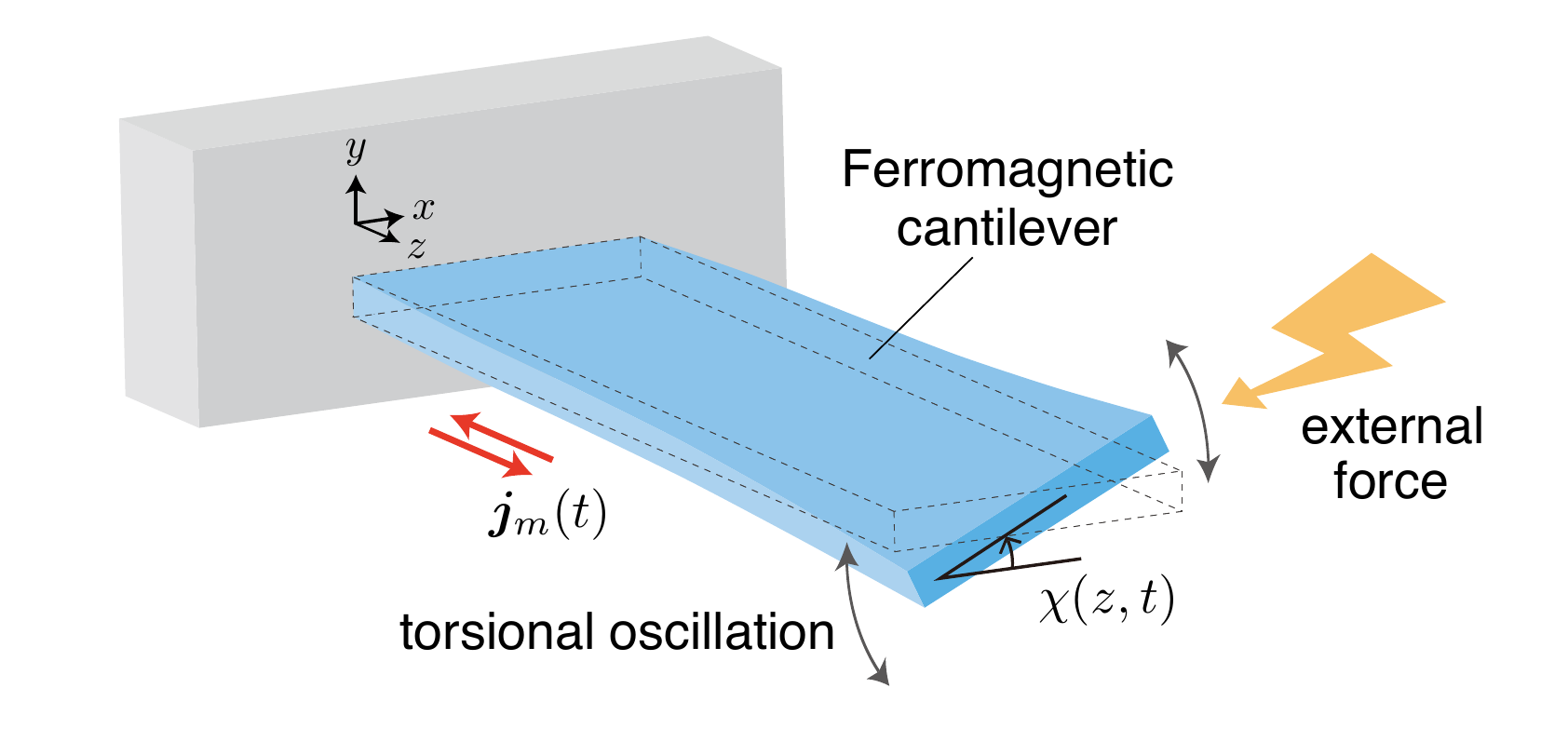}
	\caption{\label{fig:1} Schematic setup of our theory. Considering that an external force excites the torsional oscillation mode, which is described by $\chi (z, t)$, we show that the magnon current $\bm{j}_m (t)$ is induced by the torsional oscillation.
	}
\end{figure}

In this Letter, we show that the spin wave propagation is induced by torsional oscillation of a nanomechanical cantilever composed of the ferromagnetic insulator.
Figure~\ref{fig:1} depicts the schematic setup of our theory, where a torsional oscillation mode is excited by an external force, such as by piezoelectric actuator, or by laser Doppler vibrometer, which results in the spin wave propagation, or more strictly speaking, the magnon current generation with frequency same as that of the torsional oscillation.
To capture the physics, we begin with the Lagrangian of a simple localized spin system, which contains the exchange interaction and easy magnetic anisotropy, where the anisotropy direction is modulated by the torsional oscillation.
By introducing a local rotation in spin space, we move to the coordinate frame in which the easy magnetic anisotropy is constant for time and space, which leads to a kind of Dzyaloshinskii-Moriya~(DM) interaction~\cite{dzyaloshinsky1958,moriya1960} emerging in the rotated frame.
In the DM interaction, the spin gauge field~\cite{tatara2008} acts as the $D$ vector and is proportional to the spatial derivative of the torsional oscillation angle, which indicates that the torsional oscillation can be described by the spin gauge field.
Hence, we evaluate the magnon current as the linear response to the spin gauge field, which we find is large enough to be detected, such as by the inverse spin Hall effect~\cite{saitoh2006,kimura2007,sinova2015}.
A possible experimental configuration is also proposed.

 We emphasize that the present theory is essentially different from theories based on the conventional magnetoelastic coupling~\cite{ruckriegel2014,cornelissen2016}, in which only the symmetric strain tensor is considered, while we here consider the torsional oscillation which is described by the antisymmetric strain tensor~\cite{landau1986}.
In order to derive the Hamiltonian containing the antisymmetric strain tensor, an approach similar to ours has been taken by Jaafar \textit{et al.}~\cite{jaafar2009}, although it is not described by the spin gauge field and the authors do not mention the DM interaction.

\begin{figure}[tbhp]
	\centering
	\includegraphics[width=\linewidth]{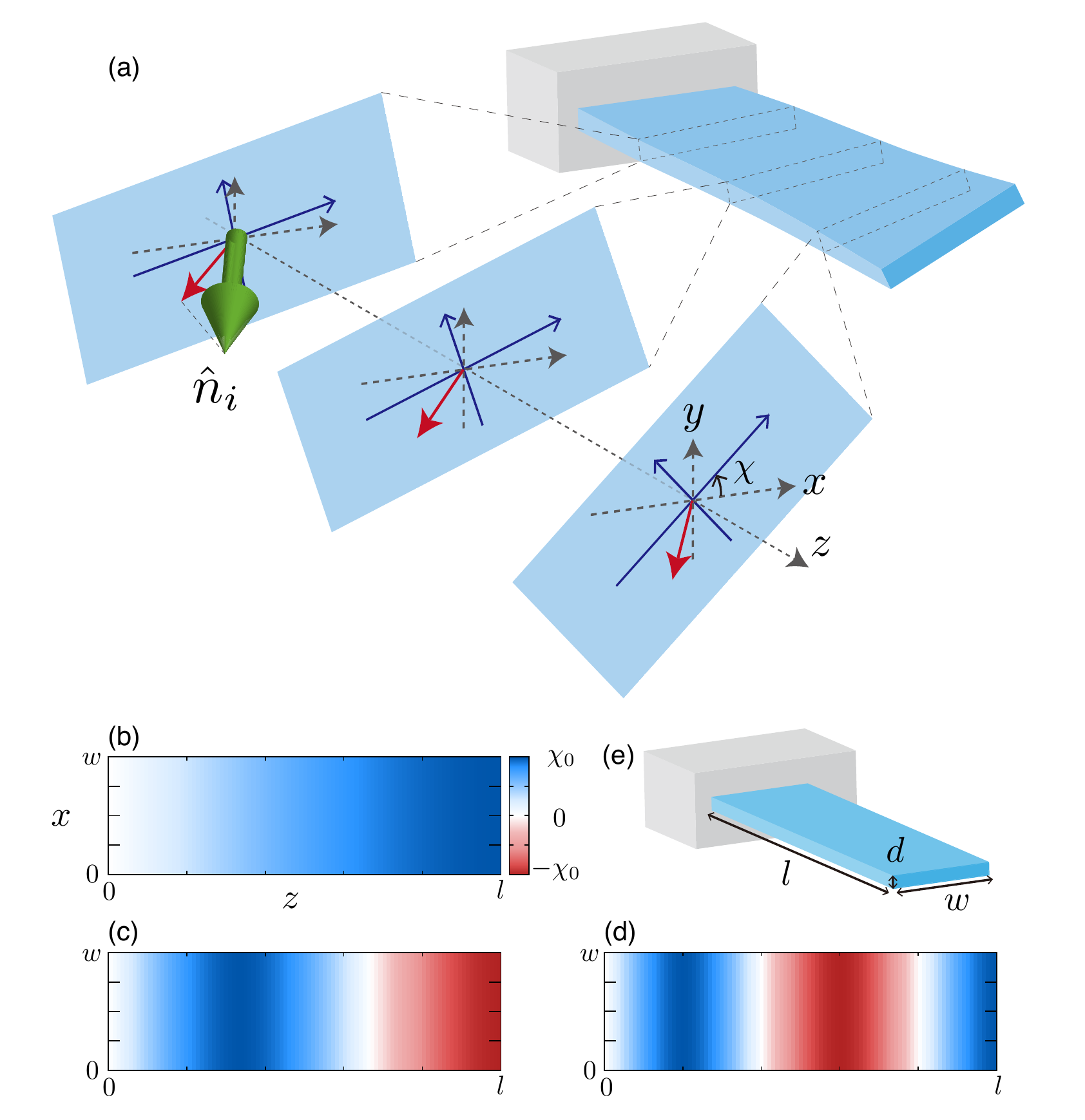}
	\caption{\label{fig:2}
	(a)~Schematic description of the magnetic anisotropy vector $\hat{n}_i$ changing by the torsional distortion.
	The gray dotted lines with arrows represent the coordinate axis in the laboratory frame, the blue lines with arrows describe the coordinate axis fixed in the cross-section of the sample, and each red arrow stands for the projection of the anisotropy vector into each cross-section of the sample.
	The red arrows do not change in the coordinate fixed in the sample, but is modulated in the laboratory frame.
	(b)--(e)~The spatial profile of the torsional oscillation angle $\chi (\br, t)$ of (b)~the lowest, (c)~second lowest, and (d)~third lowest modes at a certain time, with (e)~the schematic configuration of the sample ferromagnet.
	We can see that the torsional oscillation angle depends only on the length direction, not on the width direction.
	}
\end{figure}

Now, we begin with the localized spin Lagrangian, which is given by
\begin{align}
\mathcal{L}_s
	& = \hbar S \sum_{j = 1}^{N} ( \cos \theta_j - 1) \frac{d \phi_j}{d t} - \mathcal{H}_s
\label{def:L}
,\end{align}
where we expressed the $j$th spin $\bm{S}_j$ with length $S$ as the coherent state, $\bm{S}_j = S ( \sin \theta_j \cos \phi_j, \sin \theta_j \sin \phi_j, \cos \theta_j )$, and $N$ is the total spin number.
In this work, to reveal the essence of physics, we consider a simple situation, where the spins interact ferromagnetically through the exchange interaction with the strength $J_{\mathrm{ex}}$ and are affected by the easy magnetic anisotropy, so that the Hamiltonian $\mathcal{H}_s$ is given as
\begin{align}
\mathcal{H}_s
	& = - J_{\mathrm{ex}} \sum_{i,j} \bm{S}_i \cdot \bm{S}_j
		- \sum_i \frac{K}{2} ( \bm{S}_i \cdot \hat{n}_i )^2
,\end{align}
where $K$ is the magnitude of the anisotropy and $\hat{n}_i$ is the unit vector representing the anisotropy direction, which is temporally and spatially varying due to the torsional oscillation of the sample~[see Fig.~\ref{fig:2}~(a)].
We note that for pure torsional vibration each cross-section of the sample performs rotary vibrations about its centre of mass, which remains at rest~\cite{landau1986}; the torsional angle $\chi (\br, t)$ only depends on the $\hat{z}$ direction and time $t$.

Here, we introduce the rotational matrix $\mathcal{R}$ in order to take a frame fixed in the sample, in which the anisotropy direction is constant in time and space,
\begin{align}
\hat{n}_{i}
	& = \hat{n} (\br_i, t)
	= \mathcal{R} (\br_i, t) \hat{n}_0
\label{eq:n}
,\end{align}
where $\hat{n}_0$ is the anisotropy vector in the absence of the distortion, which is temporally and spatially constant, $\br_i$ is the position of $i$th spin, and $\mathcal{R} (\br, t)$ is given by
\begin{align}
\mathcal{R} (\br, t)
	& = \begin{pmatrix}
		\cos \chi (\br, t)
	&	- \sin \chi (\br, t)
	&	0
	\\	\sin \chi (\br, t)
	&	\cos \chi (\br, t)
	&	0
	\\	0
	&	0
	&	1
	\end{pmatrix}
,\end{align}
since we introduced the coordinate as in Fig.~\ref{fig:1}.
We assume that the torsional oscillation is driven by an external force, such as by a piezoelectric actuator, and then the distortion angle $\chi (\br, t)$ obeys the following equation of motion~\cite{landau1986},
\begin{align}
C \frac{\partial^2 \chi}{\partial z^2} = \rho I \frac{\partial^2 \chi}{\partial t^2}
\label{eq:eom}
,\end{align}
where $C$ is an elastic constant defined by the shape and material of the sample~\cite{kovalev2005}, $\rho$ is the mass density, and $I$ is the moment of inertia of the cross-section about its center of mass.
For a plate with thickness $d$ and width $w$ ($ \gg d$), the quantities $C$ and $I$ are given as $C = \mu d^3 w / 3$ and $I \simeq d w^3 / 12$, where $\mu$ is the Lam\'{e} constant~\cite{kovalev2005}.
The solution of Eq.~(\ref{eq:eom}) at a certain time is shown in Fig.~\ref{fig:2}~(b)--(d) [see Eq.~(\ref{eq:solution}) for details].

Taking the continuum limit $\bm{S}_i \to \bm{S} (\br)$, we move to the rotated frame, $\tilde{\bm{S}} = \mathcal{R}^{-1} \bm{S}$, where the Hamiltonian is obtained as $\mathcal{H}_s = \mathcal{H}_0 + \mathcal{H}_A$.
The first term $\mathcal{H}_0$ contains the corresponding terms to the exchange interaction and the easy magnetic anisotropy, $\mathcal{H}_0 = \int (d \br/a_0^3) [ (J/2) (\partial_i \tilde{\bm{S}}) \cdot (\partial_i \tilde{\bm{S}}) - (K/2) (\tilde{\bm{S}} \cdot  \hat{n}_0 )^2 ]$, where $J = 2 J_{\mathrm{ex}} a_0^2$ and $a_0$ is the lattice constant of the sample ferromagnet.
In the rotated frame, the additional term $\mathcal{H}_A$ appears;
\begin{align}
\mathcal{H}_A
	& = \int \frac{d \br}{a_0^3} \left[
		- J \bm{A}_i \cdot ( \tilde{\bm{S}} \times \partial_i \tilde{\bm{S}} )
		- \hbar \tilde{\bm{S}} \cdot \bm{A}_t
	\right]
\label{eq:H_A}
,\end{align}
which is proportional to spin gauge field $\bm{A}_{\mu} = (A^x_{\mu}, A^y_{\mu}, A^z_{\mu})$ with $\mu = t, x, y, z$.
The spin gauge field is connected to the rotational matrix as
\begin{align}
( \mathcal{R}^{-1} \partial_i \mathcal{R} )_{\alpha \beta}
	= A_i^{\gamma} \epsilon^{\alpha \beta \gamma}
\end{align}
with $i = x, y, z$~\cite{kohno2007}, so that we find $A_i^{\alpha} = (\partial_i \chi) \delta^{\alpha, z}$, and $\bm{A}_t = \dot{\chi} \hat{z}$.
We here emphasize that the first term in Hamiltonian~(\ref{eq:H_A}) is nothing but the Dzyaloshinskii-Moriya~(DM) interaction~\cite{dzyaloshinsky1958,moriya1960}, which means that torsional distortion in ferromagnets induces DM interaction.

Next, in order to use the Holstein-Primakoff~(HP) transformation, we further introduce the global rotational matrix $\mathcal{R}_0$ defined by $\hat{n}_0 = \mathcal{R}_0 \hat{z}$, and also introduce $\bar{\bm{S}} = \mathcal{R}_0^{-1} \tilde{\bm{S}}$ and $\bar{\bm{A}}_{\mu} = \mathcal{R}_0^{-1} \bm{A}_{\mu}$.
In the frame described by $\bar{\bm{S}}$, we safely use the HP expansion,
$\bar{S}^{x} (\br) \simeq \sqrt{2 S} [a (\br) + a^{\dagger} (\br)]/ 2$,
$\bar{S}^{y} (\br) \simeq \sqrt{2 S} [a (\br) - a^{\dagger} (\br)]/ 2 \zi$,
and $\bar{S}^z (\br) = S - a^{\dagger} (\br) a (\br)$.
Hence, the Hamiltonian in the Fourier space is given as $\mathcal{H}_0 = \sum_{\bq} \omega_{\bq} a^{\dagger}_{\bq} a^{}_{\bq}$ with $\omega_{\bq} = \mathcal{J} q^2 + \Delta$, where $\mathcal{J} = S J = 2 S J_{\mathrm{ex}} a_0^2$ and $\Delta = S K$~\footnote{We redefined the magnon operators as $a^{(\dagger)} (\br) \to a_0^{3/2} a^{(\dagger)} (\br)$.}, and
\begin{align}
\mathcal{H}_A
	& = \hbar \sum_{\bp} \bar{j}_i (-\bp) \bar{A}_i^z (\bp, t)
		+ \hbar S \sum_{\bp} \bar{n} (-\bp) \bar{A}_t^z (\bp, t)
\label{eq:H_A_magnon}
,\end{align}
where $\bar{\bm{j}}_m (\bp) = \hbar \bar{\bm{j}} (\bp)$ is the magnon current density operator in the frame described by $\bar{\bm{S}}$, with
\begin{align}
\bar{j}_i (\bp)
	& = \frac{2 \mathcal{J}}{\hbar V} \sum_{\bq} q_i a^{\dagger}_{\bq-\bp/2} a^{}_{\bq+\bp/2}
,\end{align}
and $\bar{n} (\bp)$ is magnon density operator given by $\bar{n} (\bp) = V^{-1} \sum_{\bq} a^{\dagger}_{\bq-\bp/2} a^{}_{\bq+\bp/2}$.
Here, $V$ is the volume of the sample ferromagnet.

\begin{figure}[btp]
	\centering
	\includegraphics[width=\linewidth]{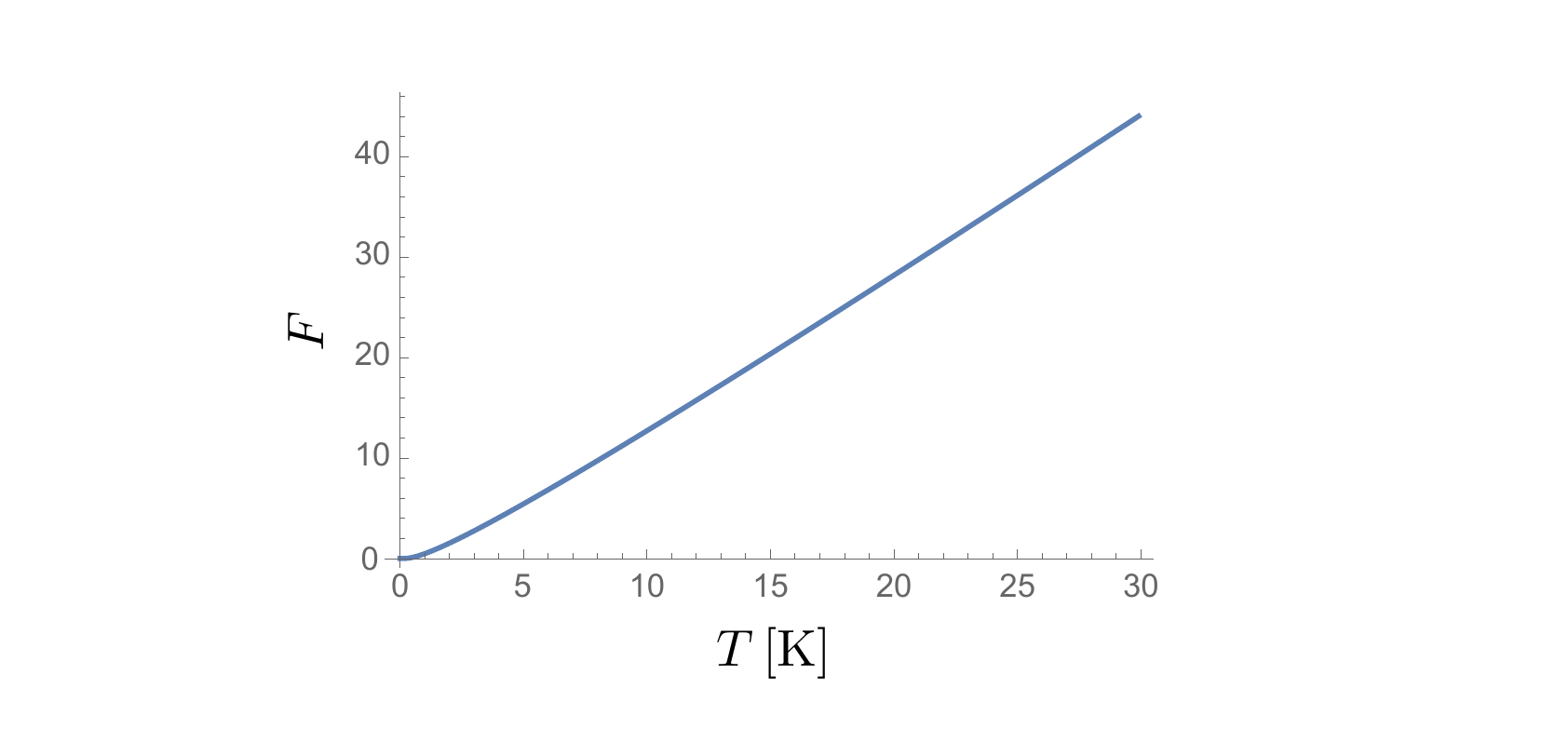}
	\caption{\label{fig:3}Temperature dependence of the function $F(\beta \Delta)$ with the gap $\Delta / \kB \simeq 0.67 \,\mathrm{K}$ for YIG~\cite{cherepanov1993}.
	}
\end{figure}

According to Eq.~(\ref{eq:H_A_magnon}), the spin gauge field couples to the magnon current density, so that we easily predict magnon current generation by the dynamical distortion.
We now evaluate the linear response of the magnon current to the torsional oscillation, which is given by
\begin{align}
\bar{j}_{m, i} (\bp, \Omega)
	& = \hbar \langle \bar{j}_i (\bp, \Omega) \rangle
	= \bar{\chi}_{i j}^{\R} (\bp, \Omega) \bar{A}_j^z (\bp, \Omega)
,\end{align}
where the response coefficient is obtained from
\begin{align}
\bar{\chi}_{i j} (\bp, \zi \omega_{\lambda})
	& = - \hbar^2 \int_0^{\beta} d\tau e^{\zi \omega_{\lambda} \tau} \langle T_{\tau} \bar{j}_i (\bp, \tau) \bar{j}_j (-\bp, 0) \rangle
\label{def:chi}
\end{align}
with $\beta = 1/\kB T$, by taking the analytical continuation, $\zi \omega_{\lambda} \to \hbar \Omega + \zi 0$~\cite{kubo1957,fujimoto2019}.
As standard procedures of the calculation for the linear response theory, rewriting Eq.~(\ref{def:chi}) by means of the thermal Green function of magnon, replacing the Matsubara summation with the contour integral, and taking the analytical continuation, we then focus on the $\Omega$-linear term in the response coefficient,
\begin{align}
\Omega \left[ \frac{\partial}{\partial \Omega} \bar{\chi}_{i j}^{\R} (\bp = 0, \Omega) \right]_{\Omega = 0}
	& = - \zi \hbar \Omega \Phi \delta_{ij}
,\end{align}
where we neglected the $\bp$ dependence of $\bar{\chi}^{\R}_{i j}$ because the spin gauge field $\bar{A}_j^z$ is already first order of $\bp$, and the lowest order is of our interest.
Here, $\Phi$ is given by
\begin{align*}
\Phi
	& = \frac{2 \mathcal{J}}{3} \frac{1}{V} \sum_{\bq} \left( \omega_{\bq} - \Delta \right)
		\int_{-\infty}^{\infty} \frac{\dd{\epsilon}}{2 \pi} \left( \frac{\partial g}{\partial \epsilon} \right)\left\{ \Im\!\left[ D^{\R}_{\bq} (\epsilon) \right] \right\}^2
,\end{align*}
where $g = (e^{\beta \epsilon} - 1)^{-1}$ is the Bose-Einstein distribution function, and $D^{\R}_{\bq} (\epsilon)$ is the retarded Green function of magnon,
\begin{align}
D^{\R}_{\bq} (\epsilon)
	& = \frac{1}{\epsilon - \omega_{\bq} + \zi \alpha \epsilon}
,\end{align}
which is obtained from the Landau-Lifshitz-Gilbert equation with the phenomenologically-introduced Gilbert damping constant $\alpha$.
By taking the approximation $\Im [D^{\R}_{\bq} (\epsilon)]^2 \simeq (2 \pi/\alpha \epsilon) \delta (\epsilon - \omega_{\bq})$, we obtain
\begin{align}
\Phi
	& = \frac{1}{6 \alpha \pi^2} \sqrt{ \frac{\Delta}{\mathcal{J}} } F (\beta \Delta)
\label{eq:Phi}
\end{align}
with $ F (x) = x^{-1/2}\int_{x}^{\infty} \dd{t} (t - x)^{3/2} e^{t} / t (e^t - 1)^2$~(see Fig.~\ref{fig:3}).
Hence, the real time and space representation of the magnon current density induced by the spin gauge field is given as $\bar{j}_{m, i} (\br, t) = \hbar \Phi \dot{\bar{A}}_i^{z} (\br, t)$ in the frame described by $\bar{\bm{S}}$, that is,
\begin{align}
j_{m, i} (\br, t)
	= \hbar \Phi [ \hat{n}_0 \cdot \dot{\bm{A}}_i (\br, t) ]
\label{eq:j_m}
\end{align}
in the rotated frame described by $\tilde{\bm{S}}$.
Equation~(\ref{eq:j_m}) with Eq.~(\ref{eq:Phi}) is the main result of this work.
The spin polarization direction of magnon current is almost parallel to $\hat{n}_0$ even in the laboratory frame, because the torsional oscillation angle is much smaller; $\chi \ll 1$; especially in the edges $\chi = 0$.
We also note that the flow direction of the magnon current is along the length direction, since $\dot{\bm{A}}_x = \dot{\bm{A}}_y = 0$ and $\dot{\bm{A}}_z = [\partial_t \partial_z \chi (z, t)] \hat{z}$.

We now estimate the magnitude of the generated magnon current.
We first determine the dynamics of the torsional oscillation, which is governed by Eq.~(\ref{eq:eom}).
Assuming the boundary conditions of $\chi (z, t)$ as $\chi (0, t) = 0$ and $\partial_z \chi (l, t) = 0$, and the initial conditions $\chi(l, t_0) = \chi_0$ and $\partial_t \chi (z, t = t_0) = 0$, we obtain
\begin{align}
\chi (z, t)
	& = \chi_0 \sin k z \cos (\Omega t + \delta)
\label{eq:solution}
\end{align}
with $k = (n - 1/2) \pi / l$ and $\Omega = v k$, where $n$ is a natural number, and $v = (2 d / w) \sqrt{\mu / \rho}$.
The phase factor $\delta$ is determined from $\cos (\Omega t_0 + \delta) = 1$ and $\sin (\Omega t_0 + \delta) = 0$.
Hence, the DM interaction for the lowest oscillation mode $n = 1$ is evaluated as $| J \bm{A}_i | / a_0 \simeq 6.7 \times 10^{-9} \,\mathrm{eV}$ for YIG of $\mathcal{J} = 5.279 \times 10^{-21} \,\mathrm{eV\,m^2}$, $a_0 = 1.2376 \,\mathrm{nm}$, and $S = 10$~\cite{cornelissen2016} with $l = 1 \,\mu \mathrm{m}$ and $\chi_0 \sim 0.01$, which is very weak compared to the other energy scales, but its time derivative is important for the magnon current generation, which is large enough to be detected because of $\Omega \simeq 4.8 \times 10^9 \,\mathrm{s}^{-1}$ for $d / w = 0.4$ with $\mu = 75.42 \times 10^9 \, \mathrm{kg/m \,s^2}$ and $\rho = 5.1 \times 10^3 \,\mathrm{kg/m^3}$~\cite{chou1988}.
Indeed, the magnon current density at the edge is calculated as converting the unit into that of the electric current as
\begin{align}
	I_z (t)
	& = j_{m, z} (0,t) \frac{2 e}{\hbar} d w
\\
	& \le \frac{2 e \chi_0}{3 \alpha} \sqrt{ \frac{\Delta}{\mathcal{J}} \frac{\mu}{\rho} } F (\beta \Delta)
		\left( n - \frac{1}{2} \right)^2
		\left( \frac{d}{l} \right)^2
		(\hat{n}_0 \cdot \hat{z})
\notag \\
	& \simeq 2 \,\mu \mathrm{A}
, \qquad \text{for $n = 1$}
\notag
.\end{align}
Here, we used $\Delta \simeq 5.8 \times 10^{-5} \mathrm{eV}$, and $\alpha \sim 10^{-4}$ for YIG~\cite{cherepanov1993}.
We also assumed $\chi_0 \sim 0.01$, $d / l = 0.2$ and $T = 30 \,\mathrm{K}$.
Note that the calculated magnon current $I_z$ depends on the ratio of the thickness and length, $d / l$, not on the width $w$, and is proportional to the square of the oscillation mode number $n$.
We further point out that although the emergent DM interaction is weak for YIG, there would be relevant phenomena in multiferroic materials, where the exchange interaction is much more strong, and the distortion scale could be comparable to the lattice constant~\cite{mochizuki2011}.


\begin{figure}[t]
	\centering
	\includegraphics[width=\linewidth]{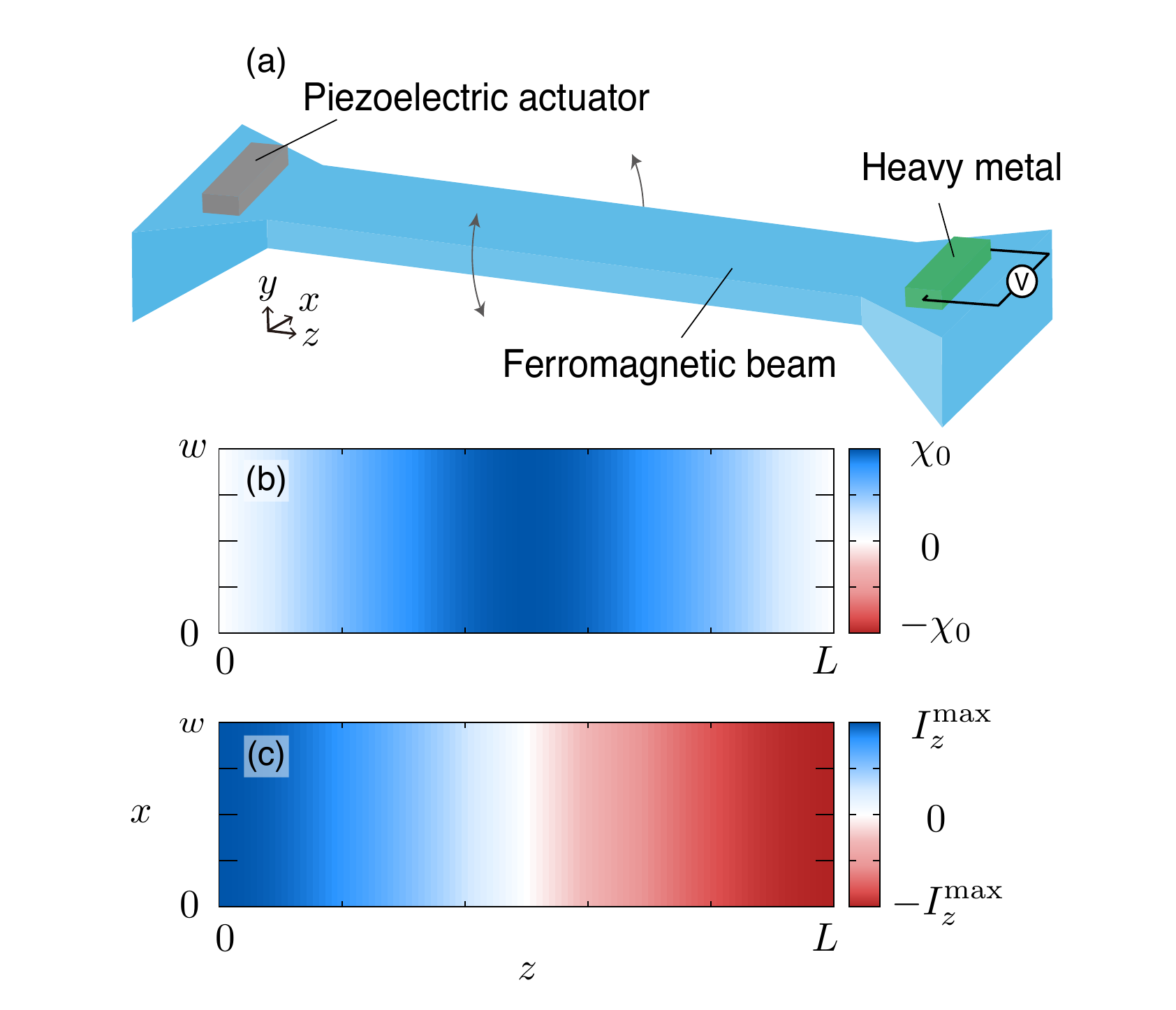}
	\caption{\label{fig:4}(a)~Schematic description of the possible experimental setup, where a nanomechanical beam structure of ferromagnetic insulator is attached by a piezoelectric actuator and by a heavy metal in which the spin Hall angle is large.
	The piezoelectric actuator excites the torsional oscillation of the beam, which induces the magnon current and then the inverse spin Hall current is detected in the heavy metal.
	(b) Spatial profile of the torsional oscillation angle, and (c) the corresponding spatial profile of the magnon current generated by the torsional oscillation, where $L$ is the length of the beam.
	We note that the magnon current takes a maximum value at the edge of the sample.}
\end{figure}
 Figure~\ref{fig:4}~(a) presents the schematics of an experimental setup to detect the magnon current generated by the torsional oscillation.
We consider a YIG nanomechanical beam structure attached by a piezoelectric actuator and by a detector composed by heavy metal with large spin Hall angle.
Figure~\ref{fig:4}~(b) depicts the spatial profile of the torsional oscillation of the lowest mode, and Fig.~\ref{fig:4}~(c) shows the corresponding magnon current generated by the torsional oscillation.
Although the torsional angle is zero at the edges, the generated magnon current takes the maximum value at the edges, since the spatial derivative contributes to the magnon current generation.
Hence, assuming $\mathrm{Pt}$ as the heavy metal, whose spin Hall angle is about $0.9 \%$~\cite{sinova2015} with the resistivity $\rho \sim 10^{-7} \, \mathrm{\Omega m}$, the inverse spin Hall current in the heavy metal is the order of $10 \,\mathrm{nA}$, or divided by the cross-section assumed as $400 \,\mathrm{nm^2}$ with length $500 \,\mathrm{nm}$, we have $1.2 \,\mu \mathrm{V}$, which is detectably large.
We note that the spin polarization of the magnon current is almost parallel to $\hat{n}_0$, and the magnitude is proportional to $\hat{n}_0 \cdot \hat{z}$.

We also note that the magnon current generation proposed here can be regarded as an extension of the Barnett effect.
The Barnett effect is originally demonstrated for the rigid body rotation, where the rotation couples to the magnetization as an effective magnetic field.
Thus, the coupling can be interpreted as a Zeeman coupling due to the rigid rotation.
In contrast, we here show that the spatially nonuniform torsional rotation couples to the localized spin via the emergent DM interaction, resulting in generating the magnon current.

Finally, we would like to comment on possible connections of our theory to strain engineering and flexible magnetoelectronics.
The main topics of current strain engineering~\cite{dai2019} and flexible magnetoelectronics~\cite{ac2008,chen2008,ota2018} are related to only the \textit{symmetric} strain tensor.
As mentioned above, torsional mechanical motions are related to the \textit{antisymmetric} strain tensor, and couples to the spin degree of freedom.
Our theory paves the way for studying spin-nanomechanical phenomena given by antisymmetric strain tensor and will contribute to developments in strain engineering and flexible magnetoelectronics with torsion.

To conclude, we have considered the nanomechanical cantilever composed of the ferromagnetic insulator, which performs torsional oscillation, and shown that the magnon current is induced by the torsional oscillation.
We find that the torsional oscillation can be described by the spin gauge field, which produces a kind of DM interaction.
From the evaluation of the linear response of the magnon current to the spin gauge field by using the Matsubara Green function method, we obtain the microscopic form of the magnon current.
The estimation of the value suggests that the magnon current is detectably large by the inverse spin Hall effect.
The possible experimental setup is also presented.
As we have seen, the spin gauge field is a powerful tool to approach the torsional spin-nanomechanical effect, and our theory opens a new avenue for studying torsional spin-nanomechanical phenomena by using the spin gauge field.

\par
We thank Y.~Nozaki, H.~Chudo, T.~Narushima, K.~Yamanoi, T.~Horaguchi, G.~Okano, and S.~Tateno for giving stimulating information.
We also thank Y.~Ominato for fruitful discussion.
This work is partially supported by the Priority Program of Chinese Academy of Sciences, Grant No. XDB28000000.

\bibliography{reference}
\end{document}